\documentclass{WileyMSP-template}
\usepackage[superscript]{cite}

\begin{document}

\pagestyle{fancy}
\rhead{\includegraphics[width=2.5cm]{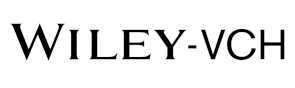}}

\sloppy

\title{Recyclable Organic Bilayer Piezoresistive Cantilever for Torque Magnetometry  at Cryogenic Temperatures}

\maketitle


\author{Eden Steven}*,
\author{Danica Krstovska}*,
\author{Daniel Suarez},
\author{Tasya Berliana},
\author{Eric Jobiliong},
\author{Eun Sang Choi}

\begin{affiliations}
E. Steven\\
Emmerich Research Center, Pluit Karang Permai 12 B9 Barat no 1 and 2, DKI Jakarta, 14450, Indonesia\\
E-mail: eden.steven@gmail.com\\
D. Krstovska\\
Faculty of Natural Sciences and Mathematics, Ss. Cyril and Methodius University, Arhimedova 3, 1000 Skopje, North Macedonia\\
E-mail:danica@pmf.ukim.mk\\
D. Suarez, E. S. Choi\\
National High Magnetic Field Laboratory and Department of Physics, 1800 East Paul Dirac Drive, Tallahassee, FL 32310, USA\\
T. Berliana\\
Department of Electrical Engineering, Universitas Kristen Krida Wacana, Tanjung Duren Raya No.4, DKI Jakarta, 11470, Indonesia\\
E. Jobiliong\\
Faculty of Science and Technology, Universitas Pelita Harapan, M.H. Thamrin Boulevard 1100 Lippo Village, Tangerang 15811, Indonesia\\

\end{affiliations}


\begin{abstract}
Flexible sensors made from organic bilayer films of molecular conductor on polymeric matrix have attracted many interest due to their simple fabrication with high potential for being scaled up, and for their high-performing multi-functionality at room temperatures. In particular, the piezoresistive property of the organic bilayer film is among one of the highest ever reported, allowing its utilization in various sensing applications. In this work, we present the study of the flexural piezoresistivity of an organic bilayer film based on $\beta-\rm (BEDT-TTF)_2I_3$ on polycarbonate matrix from room temperatures down to cryogenics temperatures. Non-trivial temperature dependent profile of the gauge factor is revealed, including enhancement of the gauge factor from $\sim 18$ at room temperatures to $\sim 48$ at 4.3 K. An organic bilayer cantilever magnetometer is developed and demonstrated to measure magnetic properties of a single crystalline organic superconductor $\kappa-\rm (BEDT-TTF)_2Cu(N(CN)_2)Br$ at temperatures down to $\sim 2.75$ K and magnetic fields up to 5 T. The high-performing bilayer devices can be fabricated in a very simple manner, and they are robust and recyclable.
\end{abstract}


\section{Introduction}

The role of electronic sensors encompasses almost every aspect of life, including but not limited to the development of smart energy saving systems, robotics, health and environmental monitors, and instrumentations for scientific discoveries. In some cases, it is advantageous for the sensors to exhibit mechanical flexibility, in addition to their intrinsic sensing properties. For example, flexible sensors based on silk are used for monitoring brain activity due to their softness and flexibility, therefore, allowing the sensors to conform to the target sensing area effectively.\cite{1} In other cases, carbon nanotube-based flexible sensors are used to develop smart electronic gloves and health monitoring technologies\cite{2,3}. Conducting polymers have also been used in consumer electronics, for example as touch sensors\cite{4,5}.

One of the classes of conducting polymers is known as the organic bilayer films (BL). This class of conducting polymer is based on charge transfers salts, a rich system of materials known for its rich physics\cite{6} and multi-functionalities\cite{7}. The bilayer system is composed of an electrically conducting active layer and an insulating layer, typically of polycarbonate\cite{8,9}, poly-lactic acid (PLA) \cite{10} polymers, and more\cite{11}. Several charge transfer salts have been demonstrated to serve as the active materials including but not limited to those that are based on BEDT-TTF\cite{8,9}, BEDO-TTF\cite{12,13}, MDF-TSF\cite{10} donor molecules with various possible acceptors such as $\rm I_3$\cite{8,9}, $\rm I_xBr_{1-x}$\cite{14}, $\rm I_xBr_{3-x}$\cite{15}. 
Depending on the synthesis protocols, multiple phases of these materials can be made by tuning its electronic properties. For example, $\alpha-\rm (BEDT-TTF)_2I_3$/polycarbonate film is semiconducting \cite{11,16} whereas $\beta-\rm (BEDT-TTF)_2I_3$/polycarbonate film is metallic\cite{9,11}; in some cases, under ideal growth conditions, $\beta-\rm (BEDT-TTF)_2I_3$/polycarbonate film also shows superconductivity\cite{17}. 

The tunability of bilayer films have been exploited for various devices. For example, the electrical resistance of $\alpha’-\rm(BEDT-TTF)_2I_xBr_{3-x}$/polycarbonate films is extremely sensitive to temperature\cite{15}, allowing resistive temperature monitoring with 0.005 $^\circ \rm C$ accuracy between room temperatures and  ~60 $^\circ \rm C$, thus capable of non-contact radiant heat sensing\cite{18}. In the $\rm(BEDO-TTF)_{2.4}I_3$/polycarbonate bilayer films, rectification and electrochromism have been observed \cite{13}, suggesting potential applications for flexible diodes and displays. High piezoresistivity has been reported for $\alpha-$ and $\beta-\rm (BEDT-TTF)_2I_3$/polycarbonate films that have a gauge factor of $\sim 10$ and $\sim 18$ at room temperatures, respectively\cite{16,19}, leading to applications for breathing and blood pulse monitoring\cite{19} and pressure sensors for extreme pressure up to 500 bar\cite{20}.

Despite the above mentioned advances, there are vast opportunities for even further device developments of new technologies that would give rise to a more efficient and eco-friendly devices. For example, studies of the piezoresistivity of organic bilayer films have been mainly focused on devices under longitudinal straining/stressing\cite{11,16,19}. Furthermore, most of the devices are developed for applications at room temperatures\cite{11,13,15}. 

Here we report the physical characterizations and device applications of the $\beta-\rm (BEDT-TTF)_2I_3$/polycarbonate, hereafter $\beta-\rm (ET)_2I_3$/PC, bilayer film under flexural deformations. The flexural characteristics are investigated from room temperatures down to cryogenic temperatures, revealing $\sim 270\%$ increase in the gauge factor at ~4.3 K compared to that at room temperatures. The enhanced electromechanical property is exploited for developing torque magnetometers based on the organic bilayer film cantilevers. The organic bilayer film cantilevers are simple to produce, robust, can be easily cut to various shapes and sizes, and recyclable. As a proof-of-concept, torque magnetometry is carried out using the organic bilayer film cantilever to study superconductivity and flux dynamics in an organic superconductor $\kappa-\rm (BEDT-TTF)_2Cu[N(CN)_2]Br$\cite{7,21} at cryogenics temperatures down to $\sim 2.75$ K and at magnetic fields up to 5 T.

\section{Results and Discussion}

\subsection{Interface robustness and recyclability of organic bilayer films}

Prior works have demonstrated that the organic bilayer films exhibit robust electrical properties and long term stability at ambient conditions\cite{18,19}. However, it was not clear whether the organic bilayer film would retain its robustness under flexural deformations. One of the possible problems for flexible sensors is the sliding/buckling of the active layer when flexed due to insufficient adhesion between the active layer and the base substrate. Irreversible damage could occur due to the sliding, buckling, or in the worst scenario, being peeled off.

To evaluate the interfacial adhesion between the active and base layer, we have carried out a 180-degree peel test (Figure 1) by subjecting the active layer to a tape adhesive. No observable damage was seen after the peel test at a peel rate of 50 mm/min and strength of $\sim 100$ N/m. A more rigorous peeling and shearing using the tape adhesive was also carried out which shows no observable damage to the active layer (Supporting Information Video S1). 
This simple fact is appreciable when we consider other types of flexible sensors where the active layer is often separately applied, glued or laminated on the flexible base substrate\cite{2}. In these cases, it is critical crucial to avoid physically disturbing the active layer or to implement additional protective measures such as additional laminates. However, such measures could prove problematic due to possible reduction in performance, interfacial mismatch and other problems. Unlike the typical flexible sensors where the active sensing layer is applied, sprayed, glued or laminated onto a flexible base \cite{2}, the organic bilayer film’s active sensing layer is grown from within the base substrate\cite{8,9}, and therefore is inherently anchored to the base substrate like as a plant is rooted in soil.

\begin{figure}
\includegraphics[width=\linewidth]{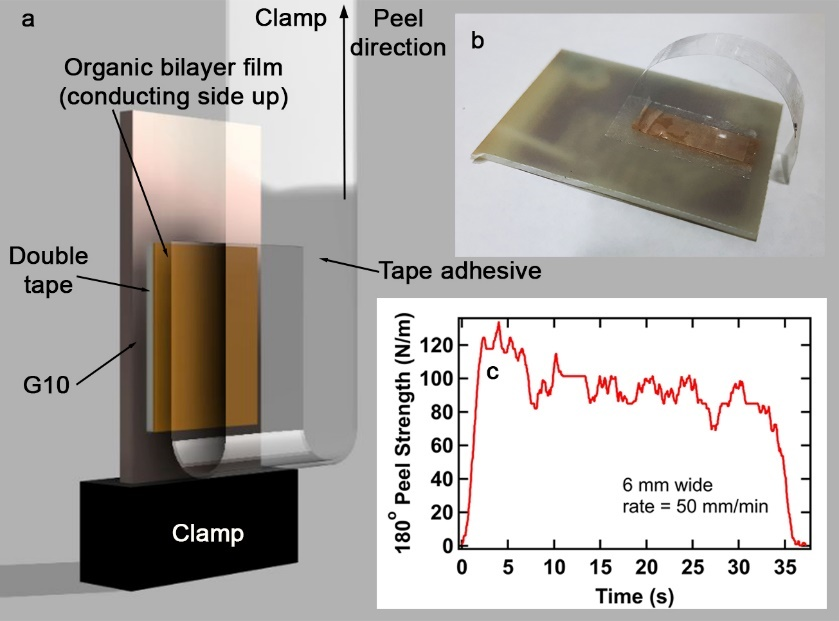}
\caption{180-degree peel test of $\beta-\rm (ET)_2I_3$/PC bilayer films. (a) Experimental setup illustration. The bilayer film is first laminated onto G10 substrate using a double sided tape. The G10 was vertically clamped at a fixed position. A layer of tape adhesive (6 mm wide) was adhered to the top surface of the bilayer film using 100 kPa pressure. (b) Photograph of the sample setup. (c) Peel strength vs time during the peel test.}
\label{fig:boat1}
\end{figure}

Organic bilayer films are fabricated through a vapor oxidation reaction of homogenously dissolved molecular crystals in a polymeric matrix. Essential to the method is the composition of the vapor which is used to dope the film. The vapor contains a solvent which acts to swell the film as well as a dopant. In the case of the $\rm (ET)_2I_3$/PC bilayer films, the base substrate is homogenously mixed ET in PC, the dopant is iodine, and dichloromethane is the swelling agent. 

As the iodine/dichloromethane reaches one of the surface of the ET/PC films, due to concentration gradient in the swelling side, ET crystals migrate and accumulate on the surface where they are then doped with the iodine. As a result, a self-assembly of $\rm (ET)_2I_3$ micro-crystallites occurs forming physically and electrically interconnected network of polycrystalline surface. After the doping process, as the swelling subsides, the micro-crystallite network is further densified forming a robust conducting active layer on one side of the film, and insulating on the other side of the film. Thus, in all of these processes, the active layer and the base substrate are always in union. 

It is also worth noting that, we have discovered that the organic bilayer film can be re-dissolved and reconstituted from the same ingredients it is formed, i.e., from the same ET and PC materials. Then, it can be re-doped into the $\beta-\rm (ET)_2I_3$/PC bilayer film repeatedly without observable degradation. We show in Figure 2, the photograph of the recycled bilayer film under an optical microscope showing similar profiles over six recycling processes. A similar order of sheet resistances of the bilayer in each process is observed. These observations suggest a simple recycling process; collecting unused, broken, torn, and mistakenly cut organic bilayer sensors and gathering them into a single vessel to be recycled into a brand new sheet of sensors for new uses, thus minimizing the hazard of such sensors to be an environmental plastic pollutant (Supporting Information Figure S1).

\begin{figure}
\includegraphics[width=\linewidth]{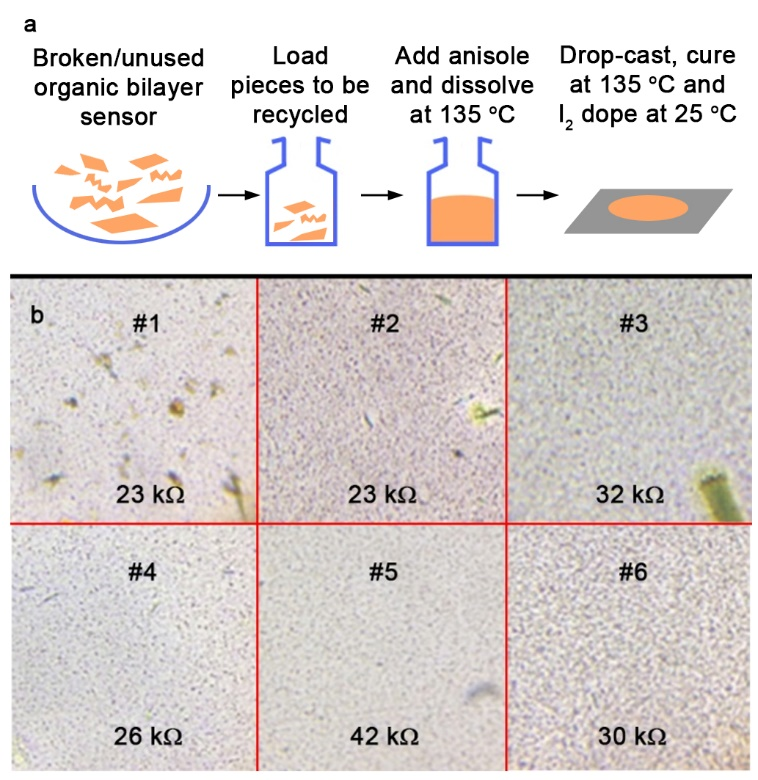}
\caption{Recycling and reconstitution of the $\beta-\rm (ET)_2I_3$/PC organic bilayer sensor. (a) Recycling process of the organic bilayer film. The dissolved solution is stable at room temperatures. Enclosed glass vial can be used for long term storage. Various shapes of film can be re-casted and re-doped depending on the applications. (b) Optical photographs of re-casted and re-doped film over 6 cycles. The films exhibit sheet resistances similar in order of magnitude.}
\label{fig:boat1}
\end{figure}

\subsection{Temperature dependent gauge factor of the $\beta-\rm (ET)_2I_3$/PC bilayer film}

The temperature dependent gauge factor of $\beta-\rm (ET)_2I_3$/PC film is obtained through temperature dependent 4-point resistance measurement at various constant flexed conditions. The full description of the method is given in the Materials and Methods section. Briefly, the film is mounted without being clamped to a linear stage by resting the film against rectilinear edges (Figure 3a). By decreasing the linear stage distance, one can induce higher flex to the film. After obtaining multiple temperature dependent resistance curves at various flex conditions, the resistance versus strain at various temperatures was obtained.

The inset in Figure 3a shows an overlay of optical photographs of the films at various flex by controlling the linear stage distance. The radius of curvature is extracted digitally (Supporting Information Figure S2) which is then used to calculate the corresponding linear strain on the film using the standard equation $\varepsilon=t/2r$, where $\varepsilon$ is the linear strain, $t$ is the thickness of the film ($\sim 25$ $\mu$m), and $r$ is the radius of curvature. In Figure 3b, the room temperature resistance corresponding to each radius of curvature is plotted, which shows reversible linear dependence during flexing and relaxing processes as expected for elastic behavior. The gauge factor is determined to be $\sim 17.5$ which is consistent with the result obtained using linear straining method reported in previous works\cite{11,19}.

\begin{figure}
\includegraphics[width=\linewidth]{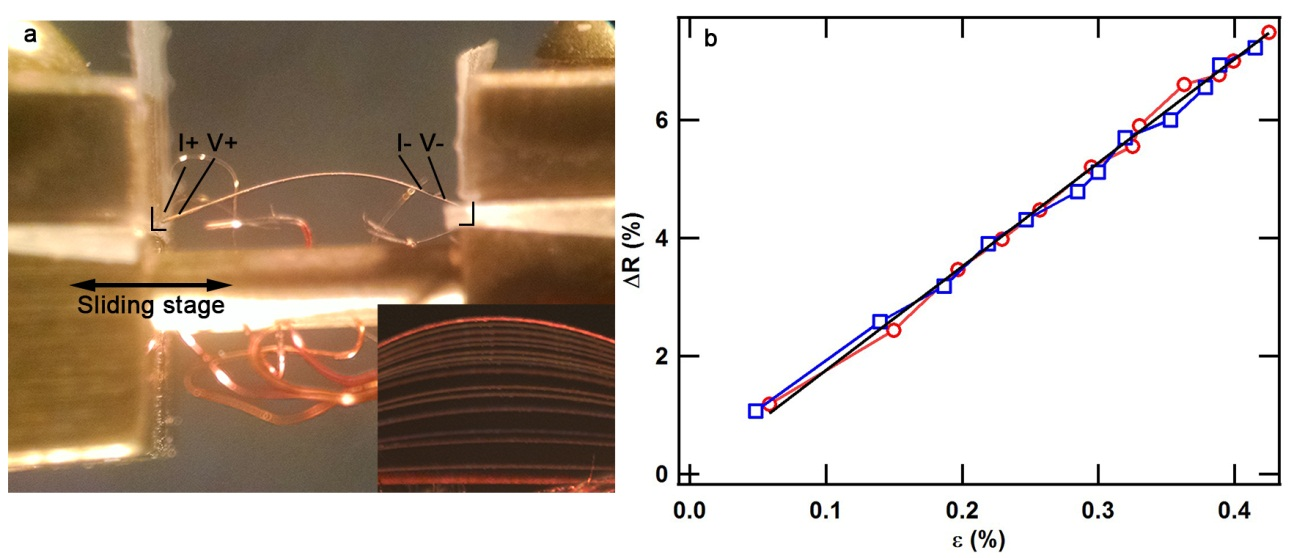}
\caption{Room temperature flexing gauge factor determination. (a) Photograph of the sample setup. Inset: side-view optical photograph of the $\beta-\rm (ET)_2I_3$/PC film under various flex. (b) Normalized resistance (against the resistance of flat film at room temperature $\sim 297$ K) vs linear strain as the film is being flexed (red circles) and relaxed (blue rectangles). Black line represents a linear fit whose slope corresponds to a gauge factor of $\sim 17.5$.}
\label{fig:boat1}
\end{figure}

There are several interesting notes concerning the benefits of determining the gauge factor of thin films via flexural deformation compared to that in typical linear straining methods. First, since the strain relates only to the thickness and radius of curvature, it is not necessary to measure the initial length of the film thus minimizing the errors that may come from inaccurate electrical contact distance measurement. Second, since the film is not clamped down (not possible for linear straining method), there is much less risk of sample damage due to over clamping or slippage due to under-clamping. Furthermore, as we cool the film down to cryogenics temperatures, the linear straining method could be affected by thermal contraction which induces internal stress to the film if not compensated.

This is especially important for the PC film since at 4.2 K it contracts up to $1.5\%$ of its initial length at room temperatures\cite{22}. If such film was clamped down on a linear stage, the internal strain induced by the cooling down process would result in an irreversible damage as the contraction would go past its elastic limit of $\sim 1\%$ strain\cite{19}. In our setup, the film is not clamped but held under natural tension as it is flexed. This allows thermal contraction of the PC film to occur freely. As the film is cooled down, the length of the film decreases which actually reduces the degree of flex of the film (i.e., increases the radius of curvature). 
In our gauge factor measurement, we have taken into account the thermal contraction effect of the film, and compensated the increasing radius of curvature during the cool down process. The full detail of the thermal expansion correction is given in the Supporting Information Section S3. 

The temperature dependent resistances of the $\beta-\rm (ET)_2I_3$/PC film at various flex are shown in Figure 4a and 4b. Seven curves are obtained and sliced to plot the resistance vs strain at various temperatures from which the gauge factor was extracted from the linear slope. Thus, the gauge factor is obtained at various temperatures and shown in Figure 4c and 4d.

\begin{figure}
\includegraphics[width=\linewidth]{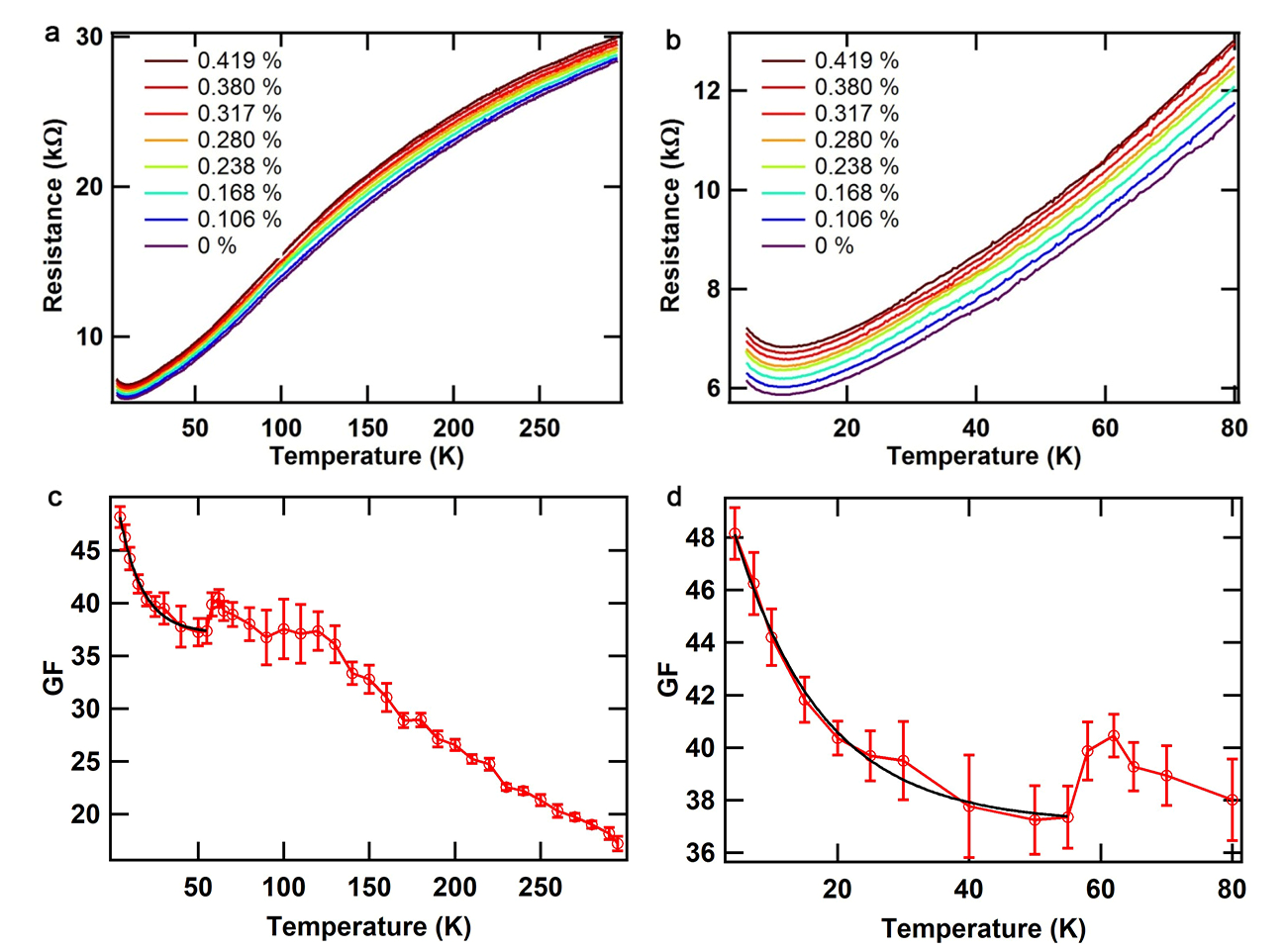}
\caption{ Temperature dependent electromechanical properties of $\beta-\rm (ET)_2I_3$. Resistance vs temperature data at various strain: (a) from room temperature and (b) from 80 K down to 4.3 K. (c, d) Temperature dependent gauge factor extracted from data in (a) and (b), respectively. Thermal contraction correction has been applied for both data shown in (c) and (d). A kink is observed at $\sim 55$ K from the temperature dependent gauge factor. We note from the resistance vs temperature data, that no observable features are seen at $\sim 55$ K. The black line represents the exponential fit line.}
\label{fig:boat1}
\end{figure}

The obtained results allow to make several interesting conclusions. First, note that at low temperatures, there is an upturn in the resistance as the sample is cooled down below $\sim 10$ K. In a perfect single crystal of $\beta-\rm (ET)_2I_3$ full metallic behavior was reported\cite{7}. The lack of the full metallic behavior in the bilayer film may suggest either slightly incomplete transformation of the $\beta-\rm (ET)_2I_3$ from the $\alpha-\rm (ET)_2I_3$ crystallites or imperfections due to grain boundaries between the micro-crystallites in the active layer. This proved to be advantageous for application of the device under consideration as described in the text below.
Second, a significant enhancement of gauge factor at lower temperatures for the $\beta-\rm (ET)_2I_3$/PC bilayer film has been revealed. Indeed, the gauge factor increases linearly between room temperature down to about 120 K, with gauge factor to temperature ratio of 0.102/K. Then it transitions into a monotonic plateau state with only a slight gauge factor increase between $\sim 120$ K to $\sim 60$ K. At $\sim 60$ K, a drop in the gauge factor is observed which is correlated with the formation of the thermally activated state; the gauge factor is exponentially increasing as the temperature is further lowered down to $\sim 4.3$ K (the lowest temperature in our measurements) according to the following equation, $\rm GF$$= y_0 + A e^{-T/B}$, where $\rm GF$ is the gauge factor, $T$ is the temperature and $y_0, A$ and $B$ are fitting constants with values of 37.125, 15.244 and 13.510 K, respectively. 
Overall, from temperatures 300 K down to 4.3 K, an incredible increase of a $\sim 270\%$ in the gauge factor of the bilayer film is observed (from $\sim 18$ up to $\sim 48$, respectively). 

\subsection{Fabrication and device setup of the organic bilayer cantilever magnetometer}

The high piezoresistivity of the $\beta-\rm (ET)_2I_3$/PC bilayer film at low temperatures opens up new venues for applications at cryogenics temperatures. Here the utilization of the bilayer film as cantilever magnetometers is elaborated. 

Conventionally, piezoresistive highly doped Si-based cantilevers are used in torque magnetometry. The gauge factor of the device is $\sim 140$. Typical arrangement consists of two levers, a reference and sensing levers, both incorporated into a Wheatstone bridge circuit. A sample under test is then glued at the end of the sensing lever where the bridge is balanced and then exposed to an external magnetic field. As the sample experiences torque and/or force by the external magnetic fields, the cantilever deflects according to the magnetic properties of the sample. The deflection is detected and recorded as differential bridge voltages. The Si-based device is highly sensitive and reliable for a wide range of measurements. However, fabrication complexities of the device are inherent in the Si-based devices. The fabrication process of such cantilever requires specialized facilities and complex lithographic processing. Also, from practical perspective, sample mounting can be a challenge, especially for a larger-sized sample. Due to limited sample mounting space, ($0.3\sim 0.5$ mm) x (0.1 mm), care must be taken in order to avoid the sample mounting adhesive from contaminating the piezoresistive element as that will impair the sensitivity. Furthermore, as in any Si-based technologies, the levers are extremely rigid and brittle, rendering the lever prone to damages due to unexpectedly large magnetic forces or torques of the sample or unintended stress during sample mounting. 

Unlike the Si-based cantilevers, the fabrication  process of the organic bilayer cantilever is very simple. As shown above (Figure 1), the robustness of the organic bilayer film allows to simplify the fabrication process to only a few low-technology steps, avoiding usually needed processes in the lithographic methods such as evaporative deposition, spin casting, washing, bonding, laminating, encapsulation and more. 

First, a piece of the organic bilayer film is generated. From this a smaller piece is cut to a desired shape using common tools such as scissors, blades or lasers, on which electrodes are finally attached by using carbon paste. In Figure 5, a setup of a prototype of the organic bilayer cantilever device is shown that is cut to a desired size by using scissors. Two pieces of the cantilever film ($\sim 1.5$ mm by $\sim 4$ mm) are prepared and mounted on a G10 substrate using PVA (polyvinyl alcohol) glue such that about 3 mm of the lever is protruding outward from the G10 base. The electrodes are established on each of the levers using carbon paste and 24 micron gold wires, after which, they are incorporated into a Wheatstone bridge configuration (Figure 5b inset). The sample under test is mounted at the tip of the sensing lever using a thin layer of silicone vacuum grease.

\begin{figure}
\includegraphics[width=\linewidth]{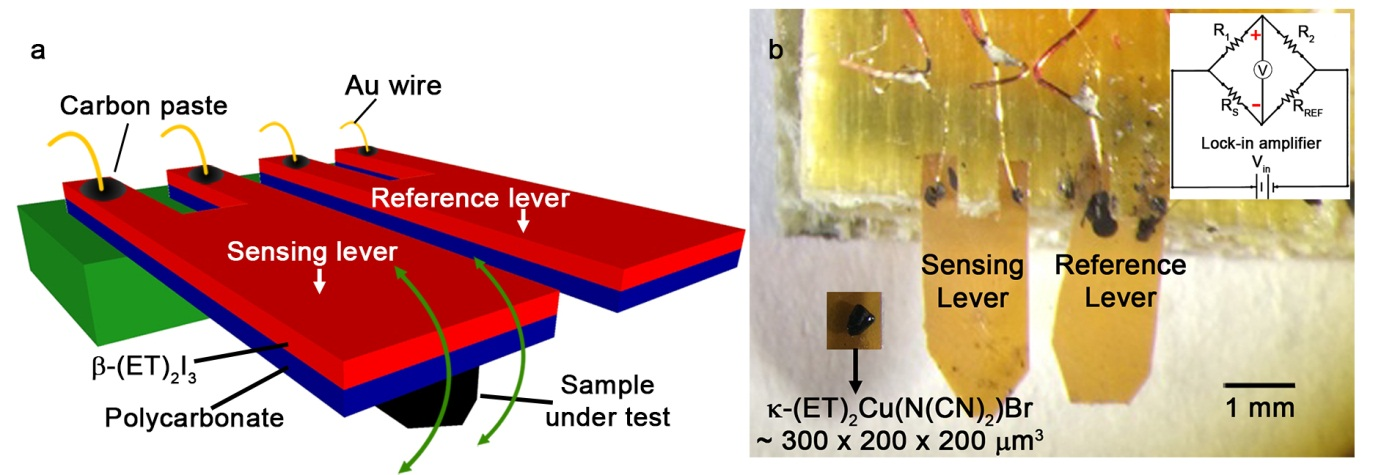}
\caption{Prototype of the $\beta-\rm(ET)_2I_3$/PC organic bilayer cantilever devices. (a) Schematic illustration of the setup. (b) Photographs of the setup. The bilayer cantilever is $\sim 1.5$ mm wide and 4 mm long, mounted on a G10 substrate using PVA (polyvinyl alcohol) glue. Inset: Wheatstone bridge circuit; RS is resistance of the sensing lever where the sample is mounted; RREF is resistance of the reference lever; R1 and R2 are variable resistances for balancing the bridge. Sample under test is $\kappa-\rm(ET)_2Cu(N(CN)_2)Br$ and it is mounted on the tip of the sensing lever using thin layer of grease.}
\label{fig:boat1}
\end{figure}

\subsection{Testing the organic bilayer cantilever magnetometer to monitor superconductivity and flux relaxation in $\kappa-\rm(ET)_2Cu(N(CN)_2)Br$}

To evaluate the performance of the organic bilayer cantilever at low temperatures and high magnetic fields, we have used a single crystal of $\kappa-\rm(ET)_2Cu(N(CN)_2)Br$ as the test sample (Figure 5b). This material is among the organic superconductors with the most robust superconductivity\cite{7,21}. In Figure 6a, the field dependent torque measurements at various temperatures are presented with field direction perpendicular to the conducting plane of the crystal. Both field up and down sweep data are obtained which show strong hysteretic torque signals that are characteristic of type II superconductors. A systematic increase in the peak signal and the critical field of the hysteresis is also observed with lowering the temperature. This behavior is expected for a superconductor as the superconductivity becomes more robust at low temperatures.

To further check whether the signal observed is coming from the sample or from the lever itself, we have also obtained the background differential signal of the Wheatstone bridge without sample attached. The result is shown in Figure 6b. It is clearly seen that the background differential signal of the levers is generally monotonic without hysteretic features. 

Additional measurement is as well performed for the purpose of confirmation that the magnetoresistance of a single lever does not show special features that may complicate sample data interpretation (Supporting Information Figure S4). However, in single crystal $\beta-\rm(ET)_2I_3$, both superconductivity ($T_c\sim 1.4$ K) and quantum oscillations are known to occur\cite{7}. In the considered bilayer film, these features are not observed most likely due to the same reasons that lead to the appearance of an upturn in the film’s resistance at low temperatures (Figure 4). Indeed, in single crystal $\beta-\rm(ET)_2I_3$, superconductivity is observed below $\sim 7$ T while above this field a beating pattern is present in the SdH quantum oscillations. The beating pattern has also been revealed in the dHvA quantum oscillations with a higher amplitude than that registered for the SdH oscillations. The fact that these properties are not present in the $\beta-\rm(ET)_2I_3$/PC film suggests that the formation of the grain boundaries in the film’s active layer induces disorder in the $\rm I^{-3}$ anions causing a change in the phonon modes responsible for the formation of the Cooper pairs. This, on the other hand, changes the prerequisites for the appearance of superconductivity and therefore it is absent in the film. Apart from that, the additional disorder significantly increases the electron scattering that may lead to a suppression of the fast magnetic quantum oscillations in the film to an extent that they cannot be experimentally resolved. Thus, the presence of a slight imperfection in the organic bilayer film turns out beneficial for the intended purpose.

The true signal from the sample is then obtained by subtracting the data shown in Figure 6b from those given in Figure 6a. The true signal is shown in Figure 6c. By comparing the raw and true signal one finds that the overall features in both signals appear to be very similar. 

\begin{figure}
\includegraphics[width=\linewidth]{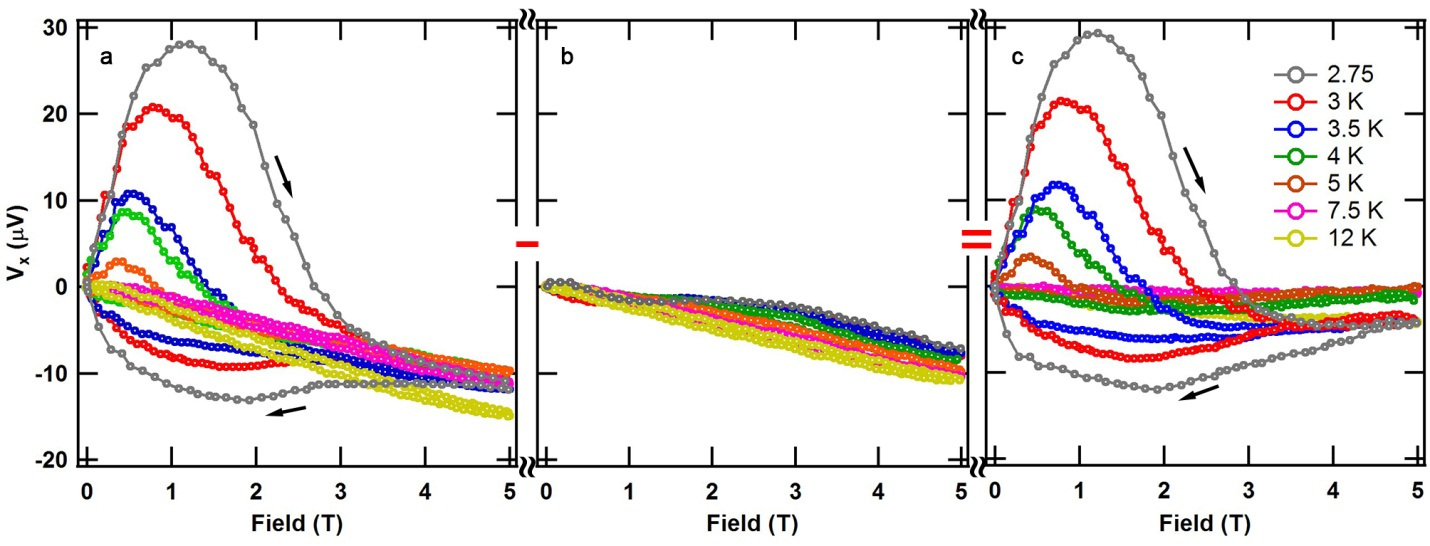}
\caption{Torque vs field sweeps of $\kappa-\rm(ET)_2Cu(N(CN)_2)Br$ single crystal at various temperatures. (a) As recorded sample plus background signals. (b) As recorded background signal. (c) Extracted sample signal after background subtraction. The loop indicates superconductivity of the sample.}
\label{fig:boat1}
\end{figure}

Most strikingly, the torque signal of the bilayer cantilever changes very smoothly at low temperatures as indicated by the absence of sharp step-like jumps in the data shown in Figure 6. Such a smooth deflection is also reflected when flux relaxation measurements were performed at various temperatures (Figure 7). The flux relaxation measurement is carried out at various constant temperatures by first ramping up the magnetic field to 1 T at a rate of 0.02 T/s after which the field is held constant and the voltage differential signal is monitored over time, corresponding to the flux relaxation of the sample. The flux relaxation follows a typical exponential relaxation behavior as evident from Figure 7. The extracted relaxation time constant $\tau$, is also confirmed to be decreasing with increasing temperatures (inset in Figure 7). This completes the validation that the organic bilayer cantilever is a viable torque magnetometer that works at cryogenics temperatures and high fields.

\begin{figure}
\includegraphics[width=\linewidth]{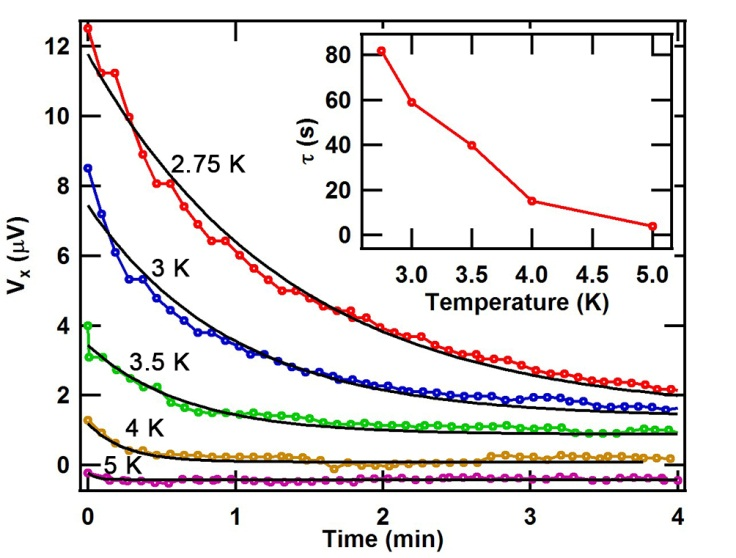}
\caption{Flux relaxation measurements of $\kappa-\rm (ET)_2Cu(N(CN)_2)Br$. The magnetic field is first swept from 0 to 1 T with a 0.02 T/s rate and then the field is held constant. The curves show the voltage differential signal while holding the field at various temperatures. The black solid line is the exponential fit to the curves. Inset: Extracted time constant $\tau$ vs temperature.}
\label{fig:boat1}
\end{figure}

\section{Conclusion}

This work highlights the simplicity, versatility, robustness, recyclability and effectiveness of the bilayer films as flexural devices. The deep anchoring of the $\beta-\rm (ET)_2I_3$ crystallites into the polycarbonate base renders the material highly compliant to flexural deformations. Combined with migration and then the densification of the crystallite network during the vapor annealing, a robust grain boundary is achieved resulting in a stable electromechanical response down to cryogenic temperatures. Electromechanical characterization of the active sensing layer down to cryogenic temperatures is carried out via iso-flexural method. The thermal contraction corrected profile reveals a non-trivial enhancement of the temperature dependent profile of the gauge factor with an observed state transition at around 60 K where the gauge factor becomes exponentially activated as the temperature is lowered. A remarkable increase in the gauge factor from $\sim 18$ to $\sim 48$ is found at temperatures of $\sim 297$ K to $\sim 4.3$ K, respectively. Using the organic bilayer film as cantilevers, magnetic torque measurements are performed on the organic superconductor $\kappa-\rm (ET)_2Cu(N(CN)_2)Br$ sample, where the signatures of superconductivity are clearly observed. Finally, this class of flexible sensors is also unique in that it can easily be re-solubilize and re-doped into pristine film without observable degradation, an aspect that is increasingly becoming important to consider.

\section{Experimental Section}

\threesubsection{$\beta-\rm (ET)_2I_3$/polycarbonate bilayer film fabrication}
BEDT-TTF (ET) crystals were homogenously dissolved at 130 $^\circ\rm C$ in polycarbonate (PC) matrix at ET:PC ratio of 2:98 using Anisole as the solvent with solid:solvent ratio of 1.5:50 (w/v), for example, 1.5 gram of solids in 50 mL solvent. The ET/PC solution was then drop casted on a glass petri dish and then dried in an oven or on a hotplate at 130 $^\circ\rm C$, for $\sim 20$ minutes until dry. Separately, a saturated $\rm I_2$ solution in dichloromethane were prepared and incubated in a glass trough with a petri dish as a cover at 25 $^\circ\rm C$ for 45 minutes to generate a stable of iodine/dichloromethane vapor. Subsequently, the petri dish cover was replaced by the petri dish containing the ET/PC dried film and the incubation at 23 $^\circ\rm C$ continued for another 3-4 minutes where the iodine/dichloromethane vapor swells the surface of the ET/PC film, causing ET crystals to migrate to the surface where they react with the iodine forming a densified polycrystalline layer of $\alpha-\rm (ET)_2I_3$\cite{11}. Then, the $\alpha-\rm (ET)_2I_3$/PC film was peeled off from the dish and annealed at 150 $^\circ\rm C$ for 15 minutes in order to allow for $\alpha-\rm (ET)_2I_3$ to convert into $\beta-\rm (ET)_2I_3$. To prevent iodine from escaping, the film was sandwiched between two glass slides during the annealing process.

\threesubsection{Temperature dependent gauge factor determination}
A cryostat equipped with a metallic linear sliding stage was used to mount the film sample. A 10 mm by 2 mm film was cut and placed resting on a rectilinear edge on the two sides of the stage without being clamped. The degree of flexing of the sample can be increased by decreasing the distance between the two ends of the stage. A four probe electrical configuration was employed via the use of carbon paste and thin 12 micron gold wires in order to ensure minimal impact of wire stiffness on the flexing of the film sample. The distance between the inner two contacts is 7 mm. Electrical resistance measurements were done using a combination of constant current source (Keithley 6221) and nanovoltmeter (Keithley 2182A). To obtain the gauge factor at various temperatures, multiple resistance vs temperature curves are obtained at various constant flex. First, the sample is flex to a desired level. Then, a scale-calibrated photograph is obtained from the side of the stage, where the radius of curvature of the flexed film is extracted using IgorPro image digitization and mathematical fittings. The sample was then cooled down from room temperature down to 4.3 K, after which the process repeats for the next constant flex conditions. In total, 12 different sets of resistance vs temperature curves are obtained at various degree of flex that corresponded to linear strain of up to $\sim 0.42$. By slicing the resistance vs temperature curves, we were able to plot and linear fit the resistance vs strain curves, obtained at various temperatures, from which the temperature dependent gauge factor was determined.

\threesubsection{Thermal contraction consideration and correction}
Thermal contraction of the polycarbonate film affects the radius of curvature as the film is cooled down. To account for this, we have included a temperature dependent correction factor for the length of the film as it has been cooled down based on the data provided by G. Schwarz\cite{22}. The thermal contraction of the film lowers the degree of flex that corresponds to an increase in the radius of curvature of the film. The details of the calculations are given in Supporting Table S1. Thermal contraction of the linear sliding stage may affect the measurement. However, since the stage was made of phosphor bronze, it has a much less pronounced thermal contraction compared to that of polycarbonate. For example, at most, cooling down to $\sim 10$ K would contract the stage by $\sim 0.3\%$ of its initial lengths at room temperature\cite{23}, whereas for polycarbonate it would contract by $\sim 1.5\%$ under similar conditions\cite{22}. Thus, the thermal contraction effects of the stage is not considered in this work. 

\threesubsection{Torque magnetometry of $\kappa-\rm (ET)_2Cu(N(CN)_2)Br$}
Two pieces of films were cut and mounted side by side on a substrate where the films were allowed to extend freely beyond the substrate to serve as levers. The two levers were setup in a Wheatstone bridge configuration, where one served as the reference lever and the other served as the sample lever. A single crystal of the organic superconductor $\kappa-\rm (ET)_2Cu(N(CN)_2)Br$ was mounted under the lever on the insulating side of the bilayer film using a minimal amount of grease. A lock-in-amplifier (Stanford Research SR830) was used to provide driving voltage and to measure the differential voltage from the bridge circuit at various magnetic fields and temperatures. The measurements were carried out in the National High Magnetic Field Laboratory (NHMFL) millikelvin facility.

\medskip
\textbf{Supporting Information} \par 
Supporting Information is available from the Wiley Online Library or from the author.

\medskip
\textbf{Acknowledgements} \par 
We thank Emmerich Research Center  and National High Magnetic Field Laboratory for the support and access to the lab facilities. A portion of this work was performed at the National High Magnetic Field Laboratory, which is supported by National Science Foundation Cooperative Agreement No. DMR-1644779 and the State of Florida. We also acknowledge NSF-DMR 1309146 for financial support.


\medskip

\begin{thebibliography}{0}

\bibitem{1}
D. -H. Kim, J. Viventi, J. J. Amsden, J. Xiao, L. Vigeland, Y. -S. Kim, J. A. Blanco, B. Panilaitis, E. S. Frechette, D. Contreras, D. L. Kaplan, F. G. Omenetto, Y. Huang, K. -C. Hwang, M. R. Zakin, B. Litt, J. A. Rogers, Nature Mater. {\bf 2010}, {\it 9}, 511.

\bibitem{2}
S. Luo, T. Liu, Adv. Mater. {\bf 2013}, {\it 25}, 565.

\bibitem{3}
E. Steven, W. R. Saleh, V. Lebedev, S. F. A. Acquah, V. Laukhin, R. G. Alamo, J. S. Brooks, Nature Commun. {\bf 2013}, {\it 4}, 2435.

\bibitem{4}
S. Ma, F. Ribeiro, K. Powell, J. Lutian, C. Møller, T. Large, J. Holbery, ACS Appl. Mater. Interfaces {\bf 2015}, {\it 7}, 21628.

\bibitem{5}
S. Takamatsu, Y. Tamashita, T. Imai, T. Itoh, Sensors and Actuators A. {\bf 2014}, {\it 220}, 153.

\bibitem{6}
K. Bechgaard, C. S. Jacobsen, K. Mortensen, H. J. Pedersen, N. Thorup, Solid State Commun. {\bf 1980}, {\it 33}, 1119.

\bibitem{7}
T. Ishiguro, K. Yamaji, G. Saito, Organic Superconductors, Springer, Heidelberg, 2nd edn, {\bf 1998}.

\bibitem{8}
J. K. Jeszka, A. Tracz, A.Polym. Adv. Technol. {\bf 1992}, {\it 3}, 139.

\bibitem{9}
E. Laukhina, C. Rovira, J. Ulanski, Synth. Met. {\bf 2001}, {\it 21}, 1407.

\bibitem{10}
A. Tracz, A. Mierczyńska, K. Takimiya, T. Otsubo, N. Niihara, J. K. Jeszka, Materials Science- Poland {\bf 2006}, {\it 24}, 517.

\bibitem{11}
E. Steven,   V. Lebedev,   E. Laukhina,   C. Rovira,   V. Laukhin,  J. S. Brooks, J. Veciana, Mater. Horiz. {\bf 2014}, {\it 1}, 522.

\bibitem{12}
H. Yamochi, T. Haneda, A. Tracz, G. Saito, Phys. Stat. Solidi B {\bf 2014}, {\it 249}, 1012.

\bibitem{13}
D. Suarez, E. Steven, E. Laukhina, A. Gomez, A. Crespi, N. Mestres, C. Rovira, E. S. Choi, J. Veciana, npj Flex Electron {\bf 2018}, {\it 2}, 29.

\bibitem{14}
E. Laukhina, V. Tkacheva, I. Chuev, E. Yagubskii, J. Vidal-Gancedo, M. Mas-Torrent, C. Rovira, J. Veciana, S. Khasanov, R. Wojciechowski, J. Ulanski, J. Phys. Chem. B {\bf2001}, {\it 105}, 11089.

\bibitem{15}
V. Lebedev, E. Laukhina, V. Laukhin, A. Somov, A. M. Baranov, C. Rovira, J. Veciana, Organic Electronics {\bf 2017}, {\it 42}, 146. 

\bibitem{16}
V. Lebedev,  E. Laukhina,  V. Laukhin,  C. Rovira,  J. Veciana, Eur. J. Inorg. Chem. {\bf 2014}, {\it 2014} 3927.

\bibitem{17}
E. E. Laukhina, V. A. Merzhanov, S. I. Pesotskii, A. G. Khomenko, E. B. Yagubskii, J. Ulanski, M. Kryszewski, J. K. Jeszka, Synt. Met. {\bf 1995}, {\it 70}, 797.

\bibitem{18}
R. Pfattner, V. Lebedev, E. Laukhina, S. Chaitanya Kumar, A. Esteban-Martin, V. Ramaiah-Badarla, M. Ebrahim-Zadeh, F. Pelayo García de Arquer, G. Konstantatos, V. Laukhin, C. Rovira, J. Veciana, Adv. Electron. Mater. {\bf 2015}, {\it 1}, 1500090.

\bibitem{19}
E. Laukhina, R. Pfattner, L. R. Ferreras, S. Galli, M. Mas-Torrent, N. Masciocchi, V. Laukhin, C. Rovira, J. Veciana, Adv. Mater. {\bf 2010}, {\it 22}, 977.

\bibitem{20}
V. Laukhin, E. Laukhina, V. Lebedev, C. Rovira, J. Veciana, Procedia Engineering. {\bf 2014}, {\it 87}, 1135.

\bibitem{21}
H. H. Wang, K. D. Carlson, U. Geiser, A. M. Kini, A. J. Schultz, J. M. Williams, L. K. Montgomery, W. K. Kwok, U. Welp, K. G. Vandervoort, S. J. Boryschuk, A. V. Strieby Crouch, J. M. Kommers, D. M. Watkins, J. E. Schriber, D. L. Overmyer, D. Jung, J. J. Novoa, M. -H. Whangbo, Synthetic Metals {\bf 1991}, {\it 42}, 1983.

\bibitem{22}
G. Schwarz, Cryogenics {\bf 1988}, {\it 28}, 248.

\bibitem{23}
A. F. Clark, Cryogenics {\bf 1968}, {\it 8}, 282. 

\end{thebibliography}
\end{document}